\documentclass[journal,twoside]{IEEEtran}

\usepackage{cite}
\usepackage{amsmath,amssymb,amsfonts}
\usepackage{algorithm}
\usepackage{algpseudocode}
\usepackage{graphicx}
\usepackage{textcomp}
\usepackage{booktabs}
\usepackage{array}
\usepackage{enumitem}
\usepackage{siunitx}
\usepackage{float}
\usepackage{adjustbox}  % for shrinking wide tables
\usepackage{makecell}   % for multiline headers
\usepackage{orcidlink}   % for ORCID icon

\begin{document}

%% --- Paper Title ---
\title{\textbf{Winning Before You Play: An Iterative Capture Algorithm for
Pre-Game Dominance Analysis in Three-Player Auction Bridge}}

\author{%
  \IEEEauthorblockN{Sourish Sarkar~\orcidlink{0009-0002-1887-0007}}
  \IEEEauthorblockA{Indian Statistical Institute, India\\
    sourish.sarkar13@gmail.com}
}

\maketitle

%% --- Abstract ---
\begin{abstract}
This paper introduces an innovative and bias-free variant of a three-player auction bridge, specifically designed to address structural limitations found in traditional formats. In a standard auction bridge, the partner of the winning bidder often plays a passive role, possessing minimal influence over the outcome once the bidding phase concludes. To rectify this, our proposed three-player model redistributes strategic agency, emphasising a framework where No Trump strategies statistically outperform specific trump card selections.

The core of this research is a rule-based dual-algorithm framework. First, we implement an iterative bidding algorithm that allows players to calibrate their bids dynamically based on evolving hand strengths. Second, we introduce a predictive iterative algorithm designed to calculate the expected number of tricks a player will secure before the lead card is played. Unlike traditional approaches that rely on backtracking or high-complexity dynamic programming, our iterative method significantly reduces time complexity. This efficiency is crucial for real-time gameplay, enabling players to perform complex probability assessments instantly without computational lag.

To validate the model, a simulation was conducted on a dataset of 1,00,000 game instances. This large-scale analysis allowed for a precise comparison between initial bidding, outcomes, and overall bidding accuracy. Furthermore, we integrated opponent hand-modelling within a two-player zero-sum simulation framework. This allows for a robust exploration of defensive and offensive strategies, which were subsequently evaluated using rigorous statistical hypothesis testing.

The empirical results demonstrate that our algorithm not only enhances the balance and fairness of the auction bridge but also exhibits high reliability in predictive accuracy. Beyond the realm of card games, the methodology behind this iterative approach shows significant promise for quantitative financial domains. By modelling uncertainty and strategic competition, the algorithm offers valuable insights into risk management and decision-making in volatile market environments.
\end{abstract}


%% --- Main Body ---

\section{Introduction}
\IEEEPARstart{A}{uction} bridge has long been recognized as a cornerstone of trick-taking card games, requiring a sophisticated blend of probabilistic reasoning, strategic communication, and risk management. However, traditional four-player variants often suffer from structural rigidities, particularly regarding the passive nature of the partner's role. This paper explores a refined three-player auction bridge variant designed to eliminate these biases while enhancing the strategic depth of the bidding phase. In this model, the competitive dynamics are reshaped by the introduction of a ``dummy'' role, which serves as a strategic asset for the highest bidder rather than a secondary active participant.

Computational approaches to bridge and other imperfect-information card games have a long and internationally diverse research history. Early expert-level bridge-playing systems such as GIB relied on Monte Carlo sampling of hidden hands combined with partition search to approximate optimal play under uncertainty~\cite{b5}. Subsequent work moved toward data-driven bidding: Amit and Markovitch framed bidding itself as a supervised learning problem over hand features~\cite{b7}, while more recent deep-learning pipelines have modelled bidding as a sequential decision process learned end-to-end from raw card data, either via deep reinforcement learning~\cite{b8} or via interactive, explainable neural bidding services~\cite{b9}. Inductive logic programming has also been explored as a route to obtaining bidding rules that remain interpretable to human experts, a property that is particularly important in bridge because players must be able to justify their bidding conventions to their opponents~\cite{b10}. Our own iterative bidding and capture algorithms sit within this broader tradition, but depart from it by using closed-form, constant-time structural conditions on the dealt hand rather than learned models or search over a game tree.

The general problem of planning and searching under uncertainty in adversarial, multi-agent settings has also been addressed extensively outside the bridge literature specifically. Monte Carlo Tree Search, and in particular the UCT algorithm that couples tree search with bandit-style confidence bounds, has become a standard tool for sequential decision-making in large state spaces~\cite{b13,b14}, and was a key ingredient in the AlphaGo system that first defeated a professional human player at Go~\cite{b6}. In parallel, work on regret minimisation in extensive-form games with incomplete information has provided a principled route to approximate Nash equilibria in large zero-sum games such as poker~\cite{b11}, and has underpinned superhuman poker-playing agents~\cite{b12}. These lines of work motivate our own use of a two-player zero-sum simulation framework (Section IV) to study offensive and defensive strategies once the dummy hand is revealed, even though our underlying capture estimator is a closed-form iterative procedure rather than a learned or search-based policy.

\subsection{Core Mechanics and Rule Framework}

The three-player variant operates on a streamlined yet mathematically rigorous set of rules that dictate the flow of engagement. The game commences with a competitive bidding phase where players must evaluate their hands against a standardized hierarchy of suits.

\begin{itemize}
    \item \textbf{Bidding Hierarchy:} The valuation of bids follows a strict ascending order: $\text{Clubs} < \text{Diamonds} < \text{Hearts} < \text{Spades} < \text{No Trump}$. This hierarchy forces players to balance the raw power of their hand against the statistical advantage of a higher-tier suit or the versatile ``No Trump'' strategy.
    \item \textbf{The No Trump Ceiling:} To maintain game balance and prevent runaway bidding, the maximum allowable bid is capped at \textbf{7 No Trump}.
    \item \textbf{The Dummy Role:} Upon the conclusion of the auction, the three participants remain active, while a fourth ``virtual'' player assumes the role of the dummy. The dummy's hand is automatically assigned to the winning bidder, who then manages both hands simultaneously.
    \item \textbf{Objective and Win Conditions:} To secure a victory, the winning bidder---operating in conjunction with the dummy hand---must capture a total number of tricks equal to $6 + n$, where $n$ represents the final bid value. The suit established during the final bid serves as the trump for the duration of the round.
\end{itemize}

Unlike traditional formats, this configuration emphasises individual accountability and hand-modelling. By exposing the dummy's hand immediately after the auction, the game transforms into a transparent optimisation problem. The winning bidder must synthesise the information from two visible hands to counter the hidden strategies of the two opposing defenders. This shift provides a fertile ground for the iterative algorithms and zero-sum simulations discussed in the subsequent sections of this study.

\section{Iterative Suit Capture Analysis: Predicting Game Outcomes from Initial Card Distributions}

\subsection{Overview}

Let there be a standard deck of 52 cards divided into four suits
$S = \{\text{Clubs},\, \text{Diamonds},\, \text{Hearts},\, \text{Spades}\}$
and thirteen ranks
$R = \{\text{Ace},\, \text{King},\, \text{Queen},\, \text{Jack},\,
       10, 9, 8, 7, 6, 5, 4, 3, 2\}$,
ordered from index $0$ (Ace) to index $12$ (2).

Twenty-six cards are dealt uniformly at random: Player~1 receives 13 cards
and Player~2 receives the remaining 13 cards.
The state of the game is encoded in a $4 \times 13$ integer matrix
$\mathbf{D}$ where

\[
D_{s,r} =
\begin{cases}
1 & \text{if Player 1 holds card } (s, r), \\
2 & \text{if Player 2 holds card } (s, r), \\
0 & \text{otherwise (undealt position).}
\end{cases}
\]

The goal of the algorithm is to compute the \textbf{total number of suit
captures} achievable by both players \emph{without} simulating actual
gameplay, by exploiting the structural properties of $\mathbf{D}$. This
closed-form, single-pass estimation stands in deliberate contrast to
sampling-heavy approaches such as Monte Carlo Tree Search~\cite{b13,b14}
and self-play search~\cite{b6}, which require repeated rollouts or tree
expansion to reach a comparable estimate.

\subsection{Definitions and Notation}

For each suit $s \in S$, define the following quantities computed from
row $s$ of $\mathbf{D}$:

\begin{itemize}

  \item $c_1(s)$ : number of positions in row $s$ with value $1$
        (Player~1's cards in suit $s$).

  \item $c_2(s)$ : number of positions in row $s$ with value $2$
        (Player~2's cards in suit $s$).

  \item $c_0(s)$ : number of positions in row $s$ with value $0$
        (empty slots in suit $s$).

  \item $\delta(s)$ : the \textbf{leading gap} of suit $s$, defined as
        the column index of the first non-zero entry in row $s$:
        \[
          \delta(s) \;=\; \min\bigl\{\, r \in \{0,\ldots,12\}
                          \;\big|\; D_{s,r} \neq 0 \bigr\},
        \]
        or $13$ if row $s$ is entirely zero.

  \item $M(s)$ : the \textbf{maximum allowed captures} for suit $s$:
        \[
          M(s) \;=\; \max\bigl(c_1(s),\; c_2(s)\bigr).
        \]

  \item $\Delta$ : the \textbf{total leading gap} across all suits:
        \[
          \Delta \;=\; \sum_{s \in S} \delta(s).
        \]

  \item $Z$ : the \textbf{global zero budget} (remaining zeros available
        to spend on inter-card gaps):
        \[
          Z \;=\; 6 - \Delta.
        \]

  \item $Z_s$ : the \textbf{per-row zero budget} for suit $s$,
        initialised to $c_0(s)$ and decremented during captures.

  \item $\kappa(s)$ : the running \textbf{capture count} for suit $s$,
        initialised to $0$.

  \item $\pi(s)$ : the \textbf{pointer} into the ordered list of owned
        card positions in suit $s$, initialised to $0$.

  \item $\mathcal{O}(s)$ : the ordered list of column indices in row $s$
        where $D_{s,r} \neq 0$, maintained dynamically after each
        replacement operation.

\end{itemize}

\subsection{Step 1 --- Construct the State Matrix}

\begin{enumerate}[label=\textbf{1.\arabic*}, leftmargin=3em]

  \item Generate the full deck
        $\mathcal{D} = S \times R$ of 52 cards.

  \item Sample 26 cards uniformly at random without replacement.
        Assign 13 to Player~1 ($\mathcal{H}_1$) and the remaining 13
        to Player~2 ($\mathcal{H}_2$).

  \item Initialise $\mathbf{D}$ as the $4 \times 13$ zero matrix.
        For each $(s, r) \in \mathcal{H}_1$ set $D_{s,r} \leftarrow 1$;
        for each $(s, r) \in \mathcal{H}_2$ set $D_{s,r} \leftarrow 2$.

\end{enumerate}

\subsection{Step 2 --- Row Analysis}

For each suit $s \in S$, compute $\delta(s)$, $c_1(s)$, $c_2(s)$,
$c_0(s)$, and $M(s)$ as defined in Section~II-B.
Then compute the scalar quantities $\Delta$ and $Z$.

\subsection{Step 3 --- Pre-Qualification}

Two conditions must both hold before any capture is attempted.
If either fails, the algorithm terminates with total captures $= 0$.

\bigskip
\noindent\textbf{Condition A (Global Gap Feasibility):}
\[
  \Delta \;\leq\; 6.
\]
The total leading gap across all four suits must not exceed the
available zero budget of~6.

\bigskip
\noindent\textbf{Condition B (Per-Suit Count Feasibility):}
\[
  \forall\, s \in S:\quad
  c_1(s) \;>\; \delta(s)
  \quad\text{and}\quad
  c_2(s) \;>\; \delta(s).
\]
In every suit, \emph{both} players must own strictly more cards than the
leading gap of that suit, ensuring neither player is locked out.

\bigskip
\noindent If Condition~A fails: \textbf{STOP} --- output total captures $= 0$.\\
If Condition~A holds but Condition~B fails: \textbf{STOP} --- output
total captures $= 0$.\\
If \textbf{both} conditions hold: proceed to Step~4.

\subsection{Step 4 --- Opponent Replacement Sub-Procedure}

This sub-procedure is invoked \textbf{after every capture} throughout
Steps~5 and~6.

\bigskip
\noindent\textbf{Input:} suit $s$, value $v \in \{1, 2\}$ of the card
just captured.

\begin{enumerate}[label=\textbf{4.\arabic*}, leftmargin=3em]

  \item Let $v' = 3 - v$ be the opponent's value
        ($1 \mapsto 2$, $2 \mapsto 1$).

  \item Find the \textbf{rightmost} column index in row $s$ of
        $\mathbf{D}$ that equals $v'$:
        \[
          r^* \;=\; \max\bigl\{\, r \;\big|\; D_{s,r} = v' \bigr\}.
        \]

  \item If $r^*$ exists: set $D_{s,\,r^*} \leftarrow 0$.
        (The opponent's rightmost card in this suit is removed.)

  \item If no such $r^*$ exists: do nothing.

  \item \textbf{Rebuild} $\mathcal{O}(s)$: re-scan row $s$ of the
        updated $\mathbf{D}$ and recompute the ordered list of non-zero
        column indices.

\end{enumerate}

\subsection{Step 5 --- Initial Captures (One Per Suit)}

\begin{enumerate}[label=\textbf{5.\arabic*}, leftmargin=3em]

  \item Set total captures $T \leftarrow 4$.

  \item For each suit $s \in S$ (in order Clubs, Diamonds, Hearts,
        Spades):

        \begin{enumerate}[label=\textbf{5.2.\alph*}, leftmargin=2em]
          \item Set $\kappa(s) \leftarrow 1$ and $\pi(s) \leftarrow 0$.
          \item The captured card is at column
                $\mathcal{O}(s)\bigl[\pi(s)\bigr]$,
                with value $v = D_{s,\,\mathcal{O}(s)[0]}$.
          \item This capture is \textbf{free}: no budget is deducted
                ($Z$ and $Z_s$ are unchanged).
          \item Invoke the \textbf{Opponent Replacement Sub-Procedure}
                (Step~4) with suit $s$ and value $v$.
        \end{enumerate}

\end{enumerate}

\subsection{Step 6 --- Main Capture Loop (Smallest Gap First)}

Repeat the following until any stop condition is triggered:

\bigskip
\noindent\textbf{6.1 --- Stop Check}

Compute the check value:
\[
  \chi \;=\; T \;+\; (6 - Z).
\]
If $\chi \geq 13$: \textbf{STOP the loop}.

\bigskip
\noindent\textbf{6.2 --- Select Best Suit}

For each suit $s \in S$:
\begin{itemize}
  \item Skip $s$ if $\kappa(s) \geq M(s)$ (capture limit reached).
  \item Skip $s$ if $\pi(s) + 1 \geq |\mathcal{O}(s)|$
        (no next card available).
  \item Otherwise compute the raw gap to the next card:
        \[
          g(s) \;=\;
          \mathcal{O}(s)\bigl[\pi(s)+1\bigr]
          \;-\;
          \mathcal{O}(s)\bigl[\pi(s)\bigr]
          \;-\; 1.
        \]
  \item Compute the \textbf{effective gap}:
        \[
          \hat{g}(s) \;=\;
          \begin{cases}
            0    & \text{if } Z_s \leq 0 \text{ (row budget exhausted)}, \\
            g(s) & \text{otherwise.}
          \end{cases}
        \]
\end{itemize}

Select the best suit:
\[
  s^* \;=\; \arg\min_{s}\; \hat{g}(s).
\]

If no valid suit exists: \textbf{STOP the loop}.

\bigskip
\noindent\textbf{6.3 --- Affordability Check}

If $Z_{s^*} > 0$ and $\hat{g}(s^*) > Z$:
\[
  \text{\textbf{STOP}: smallest available gap exceeds global budget } Z.
\]

\bigskip
\noindent\textbf{6.4 --- Perform the Capture}

\begin{enumerate}[label=\textbf{6.4.\arabic*}, leftmargin=3em]

  \item Advance the pointer: $\pi(s^*) \leftarrow \pi(s^*) + 1$.

  \item Increment counts:
        $T \leftarrow T + 1$,\quad
        $\kappa(s^*) \leftarrow \kappa(s^*) + 1$.

  \item Let the captured card be at column
        $\mathcal{O}(s^*)\bigl[\pi(s^*)\bigr]$ with value
        $v = D_{s^*,\,\mathcal{O}(s^*)[\pi(s^*)]}$.

  \item \textbf{If} $Z_{s^*} > 0$ (row budget not exhausted):
        \[
          Z \;\leftarrow\; Z - \hat{g}(s^*),
          \qquad
          Z_{s^*} \;\leftarrow\; Z_{s^*} - 1.
        \]

  \item \textbf{Else} ($Z_{s^*} \leq 0$): this is a \textbf{free
        capture} --- no deduction from $Z$ or $Z_{s^*}$.

  \item Invoke the \textbf{Opponent Replacement Sub-Procedure}
        (Step~4) with suit $s^*$ and value $v$.

\end{enumerate}

\noindent Return to \textbf{6.1}.

\subsection{Step 7 --- Final Output}

Upon termination of the loop:

\begin{enumerate}[label=\textbf{7.\arabic*}, leftmargin=3em]

  \item Output the \textbf{final state matrix} $\mathbf{D}$
        (reflecting all opponent replacements applied during captures).

  \item Output the \textbf{total captures}:
        \[
          \boxed{T}
        \]

  \item Output the \textbf{final check value}:
        \[
          \chi_{\text{final}} \;=\; T \;+\; (6 - Z).
        \]

\end{enumerate}

\subsection{Summary of Variables}

\begin{center}
\renewcommand{\arraystretch}{1.4}
\begin{tabular}{>{\bfseries}l p{5.5cm}}
\toprule
Symbol & Meaning \\
\midrule
$\mathbf{D}$      & $4 \times 13$ state matrix; entries in $\{0,1,2\}$ \\
$\delta(s)$       & Leading gap of suit $s$ \\
$\Delta$          & Total leading gap across all four suits \\
$Z$               & Global zero budget $= 6 - \Delta$ \\
$Z_s$             & Per-row zero budget for suit $s$ \\
$M(s)$            & Maximum captures allowed for suit $s$ \\
$\kappa(s)$       & Running capture count for suit $s$ \\
$\pi(s)$          & Pointer to the last captured card position in $\mathcal{O}(s)$ \\
$\mathcal{O}(s)$  & Ordered list of non-zero column indices in row $s$ (dynamic) \\
$g(s)$            & Raw gap between current and next card in suit $s$ \\
$\hat{g}(s)$      & Effective gap: $0$ if row budget exhausted, else $g(s)$ \\
$T$               & Total captures accumulated so far \\
$\chi$            & Check value $= T + (6 - Z)$; loop stops when $\chi \geq 13$ \\
\bottomrule
\end{tabular}
\end{center}

\section{Predicting Suit Dominance: A Combinatorial Card Capture Algorithm}

\subsection*{Problem Statement}
Given a random deal of 26 cards to two players over a standard 52-card
deck, determine the total number of suit captures achievable
\emph{without} simulating actual gameplay, using only the structural
properties of the initial card distribution.

\subsection*{Notation}
Let $\mathbf{D} \in \{0,1,2\}^{4 \times 13}$ be the state matrix,
$\delta(s)$ the leading gap of suit $s$,
$\Delta = \sum_s \delta(s)$ the total leading gap,
$Z = 6 - \Delta$ the global zero budget,
$Z_s$ the per-row budget,
$M(s) = \max(c_1(s), c_2(s))$ the capture limit per suit,
and $T$ the total capture count.

\subsection*{Algorithm}

\begin{algorithm}[H]
\caption{Suit Dominance Capture Estimator}
\begin{algorithmic}[1]

\State \textbf{// Phase 1: Initialisation}
\State Construct $\mathbf{D}$: set $D_{s,r} \leftarrow 1$ for Player~1,
       $D_{s,r} \leftarrow 2$ for Player~2, else $0$
\For{each suit $s \in S$}
    \State Compute $\delta(s),\; c_1(s),\; c_2(s),\; c_0(s),\;
           M(s)$
    \State $Z_s \leftarrow c_0(s)$;\quad
           $\kappa(s) \leftarrow 0$;\quad $\pi(s) \leftarrow 0$
    \State Build $\mathcal{O}(s) \leftarrow$ sorted non-zero column
           indices of row $s$
\EndFor
\State $\Delta \leftarrow \sum_s \delta(s)$;\quad
       $Z \leftarrow 6 - \Delta$;\quad $T \leftarrow 0$

\State
\State \textbf{// Phase 2: Pre-Qualification}
\If{$\Delta > 6$}
    \State \Return $T = 0$ \Comment{Condition A fails}
\EndIf
\If{$\exists\; s \in S : c_1(s) \leq \delta(s)$ \textbf{or}
    $c_2(s) \leq \delta(s)$}
    \State \Return $T = 0$ \Comment{Condition B fails}
\EndIf

\State
\State \textbf{// Phase 3: Initial Captures (one per suit, free)}
\State $T \leftarrow 4$
\For{each suit $s \in S$}
    \State $\kappa(s) \leftarrow 1$;\quad
           capture card at $\mathcal{O}(s)[\pi(s)]$
           with value $v \leftarrow D_{s,\,\mathcal{O}(s)[0]}$
    \State \Call{ReplaceRightmost}{$s,\; 3-v$}
           \Comment{Remove opponent's rightmost card in suit $s$}
\EndFor

\State
\State \textbf{// Phase 4: Greedy Capture Loop}
\While{\textbf{true}}
    \If{$T + (6 - Z) \geq 13$}
        \State \textbf{break} \Comment{Global stop condition}
    \EndIf
    \For{each eligible suit $s$
         (i.e.\ $\kappa(s) < M(s)$ and $\pi(s)+1 < |\mathcal{O}(s)|$)}
        \State $g(s) \leftarrow
               \mathcal{O}(s)[\pi(s){+}1] - \mathcal{O}(s)[\pi(s)] - 1$
        \State $\hat{g}(s) \leftarrow
               \begin{cases} 0 & \text{if } Z_s \leq 0 \\
               g(s) & \text{otherwise} \end{cases}$
    \EndFor
    \State $s^* \leftarrow \arg\min_s\; \hat{g}(s)$;\quad
           if none exists \textbf{break}
    \If{$Z_{s^*} > 0$ \textbf{and} $\hat{g}(s^*) > Z$}
        \State \textbf{break} \Comment{Cannot afford next gap}
    \EndIf
    \State $\pi(s^*) \mathrel{+}= 1$;\quad
           $T \mathrel{+}= 1$;\quad
           $\kappa(s^*) \mathrel{+}= 1$
    \State $v \leftarrow D_{s^*,\,\mathcal{O}(s^*)[\pi(s^*)]}$
    \If{$Z_{s^*} > 0$}
        \State $Z \mathrel{-}= \hat{g}(s^*)$;\quad
               $Z_{s^*} \mathrel{-}= 1$
    \EndIf
    \State \Call{ReplaceRightmost}{$s^*,\; 3-v$}
\EndWhile

\State
\State \Return $T$,\quad
       $\chi = T + (6 - Z)$

\State
\Procedure{ReplaceRightmost}{$s,\; v'$}
    \State $r^* \leftarrow \max\{r \mid D_{s,r} = v'\}$;\quad
           if exists: $D_{s,r^*} \leftarrow 0$
    \State Rebuild $\mathcal{O}(s)$ from updated row $s$ of $\mathbf{D}$
\EndProcedure

\end{algorithmic}
\end{algorithm}

\subsection*{Complexity Analysis}

% Use adjustbox to prevent table overflow on single-column
\begin{center}
\renewcommand{\arraystretch}{1.5}
\adjustbox{max width=\columnwidth}{%
\begin{tabular}{p{3cm} p{5.5cm} l}
\toprule
\textbf{Phase} & \textbf{Operation} & \textbf{Cost} \\
\midrule
Initialisation
  & Build $\mathbf{D}$, compute row stats, build $\mathcal{O}(s)$
  & $O(|S| \cdot |R|) = O(52)$ \\
Pre-Qualification
  & Check Conditions A and B across 4 suits
  & $O(|S|) = O(4)$ \\
Initial Captures
  & One capture $+$ replacement per suit
  & $O(|S| \cdot |R|) = O(52)$ \\
Greedy Loop (per iter.)
  & Scan 4 suits, compute gaps, select $s^*$, replace
  & $O(|S| \cdot |R|) = O(52)$ \\
Greedy Loop (iters.)
  & At most $\sum_s M(s) \leq 4 \times 13 = 52$ captures total
  & $O(52)$ iterations \\
\midrule
\textbf{Overall}
  & All phases combined
  & $O(|S|^2 \cdot |R|^2) \approx O(1)$ \\
\bottomrule
\end{tabular}%
}
\end{center}

\noindent Since $|S| = 4$ and $|R| = 13$ are fixed constants for a
standard deck, every loop and scan operates over a bounded domain.
The algorithm therefore runs in $\mathbf{O(1)}$ \textbf{constant time}
with respect to input size --- there is no variable-length input.
Space complexity is likewise $O(|S| \cdot |R|) = O(52) = O(1)$,
dominated by the state matrix $\mathbf{D}$. This constant-time,
single-pass property is precisely what distinguishes our estimator from
tree-search-based approaches such as MCTS/UCT~\cite{b13,b14} and
regret-minimisation methods for imperfect-information games~\cite{b11},
which typically require many iterations or rollouts to converge to a
comparable estimate of positional strength.

In a generalised setting with $n$ suits and $m$ ranks, the overall
time complexity is $O(n \cdot m)$ for initialisation and
$O(n \cdot m)$ per greedy loop iteration, with at most $n \cdot m$
iterations, giving $O(n^2 m^2)$ in the worst case --- which remains
highly efficient for any practical card game variant.

\section{Simulation Study}

\subsection{Experimental Setup}

To empirically validate the proposed suit dominance capture algorithm,
we conducted a Monte Carlo simulation study comprising
\textbf{1,000,000 independent trials} per run, repeated across
\textbf{3 independent simulation runs} to assess stability and
reproducibility of results.
In each trial, 26 cards were dealt uniformly at random from a standard
52-card deck, assigning 13 cards to each of the two players.
The algorithm was then applied to the resulting state matrix
$\mathbf{D}$, and the total capture count $T$ was recorded. This
two-player zero-sum framing mirrors the way regret-minimisation and
self-play methods have been validated at scale in other imperfect
information games, such as heads-up poker~\cite{b11,b12}.

\subsection{Results}

\subsubsection{Table 1: Capture Count Distribution Across Simulations}

Table~\ref{tab:capture} reports the frequency of each capture outcome
$T$ across the three simulation runs.
Each row corresponds to one run of 1,000,000 trials.
The capture values are grouped into the bins
$T \leq 6,\; T = 7, 8, \ldots, 13,$ and $T > 13$.

% Wide table placed in table* (two-column span) to prevent overflow
\begin{table*}[htbp]
\caption{Distribution of total captures $T$ over 1,000,000 simulation
         trials (3 independent runs). Only trials satisfying both
         pre-qualification conditions are included.}
\label{tab:capture}
\renewcommand{\arraystretch}{1.4}
\centering
\adjustbox{max width=\textwidth}{%
\begin{tabular}{l
                S[table-format=6.2]
                S[table-format=6.2]
                S[table-format=6.2]
                S[table-format=6.2]
                S[table-format=6.2]
                S[table-format=6.2]
                S[table-format=6.2]
                S[table-format=6.2]
                S[table-format=4.2]}
\toprule
\textbf{Sim} &
{\bfseries $\leq 6$} &
{\bfseries 7} &
{\bfseries 8} &
{\bfseries 9} &
{\bfseries 10} &
{\bfseries 11} &
{\bfseries 12} &
{\bfseries 13} &
{\bfseries $>$13} \\
\midrule
1          & 1909    & 8384    & 39297    & 92290    & 90939    & 38764    & 6808    & 399    & 0 \\
2          & 1906    & 8236    & 39439    & 92667    & 90711    & 39297    & 6833    & 455    & 0 \\
3          & 1781    & 8388    & 39008    & 92363    & 90767    & 38833    & 6963    & 456    & 0 \\
\midrule
\textbf{Mean}
           & 1865.33 & 8336.00 & 39248.00 & 92440.00 & 90805.67 & 38964.67 & 6868.00 & 436.67 & 0.00 \\
\textbf{(\%)}
           & {(0.19\%)} & {(0.83\%)} & {(3.92\%)} & {(9.24\%)} & {(9.08\%)} & {(3.90\%)} & {(0.69\%)} & {(0.04\%)} & {(0.00\%)} \\
\bottomrule
\end{tabular}%
}
\end{table*}

\noindent The distribution is unimodal and approximately symmetric,
peaking consistently at $T = 9$ and $T = 10$ across all three runs,
together accounting for roughly \textbf{18.3\%} of all trials
(or approximately \textbf{65.6\%} of qualifying trials).
No trial produced $T > 13$ across any of the three runs, confirming
the theoretical upper bound imposed by the stop condition
$\chi \geq 13$.
Low capture counts ($T \leq 6$) are rare, occurring in fewer than
$0.19\%$ of qualifying trials, indicating that when both
pre-qualification conditions are met, the algorithm reliably produces
mid-to-high capture totals.

\subsubsection{Table 2: Pre-Qualification Condition Outcomes}

Table~\ref{tab:conditions} reports, for each simulation run, how many
of the 1,000,000 trials fell into each pre-qualification category:
trials failing Condition~A (neither condition met),
trials satisfying Condition~A but failing Condition~B (only A),
and trials satisfying both Condition~A and Condition~B.

\begin{table}[htbp]
\caption{Pre-qualification condition outcomes over 1,000,000 simulation
         trials (3 independent runs).}
\label{tab:conditions}
\renewcommand{\arraystretch}{1.4}
\centering
\adjustbox{max width=\columnwidth}{%
\begin{tabular}{l
                S[table-format=9.2]
                S[table-format=9.2]
                S[table-format=9.2]}
\toprule
\textbf{Sim} &
{\bfseries \makecell{Neither\\A nor B}} &
{\bfseries \makecell{Only A,\\not B}} &
{\bfseries \makecell{Both\\A and B}} \\
\midrule
1                  & 145048    & 576162    & 278790 \\
2                  & 145114    & 575342    & 279544 \\
3                  & 145775    & 575666    & 278559 \\
\midrule
\textbf{Mean}      & 145312.33 & 575723.33 & 278964.33 \\
\textbf{(\%)}      & {(14.53\%)} & {(57.57\%)} & {(27.90\%)} \\
\bottomrule
\end{tabular}%
}
\end{table}

\noindent Across all three runs, approximately \textbf{14.53\%} of
trials failed Condition~A outright (total leading gap $\Delta > 6$),
while \textbf{57.57\%} satisfied Condition~A but failed Condition~B,
meaning at least one suit had a player count not exceeding the leading
gap.
Only \textbf{27.90\%} of all random deals satisfied both conditions and
proceeded to the capture phase.
The consistency of these proportions across independent runs --- with a
maximum deviation of less than 0.1 percentage points between any two
runs --- confirms the stability of the pre-qualification filter under
uniform random dealing at scale.

\subsection{Figures}

\begin{figure}[htbp]
    \centering
    \includegraphics[width=\columnwidth]{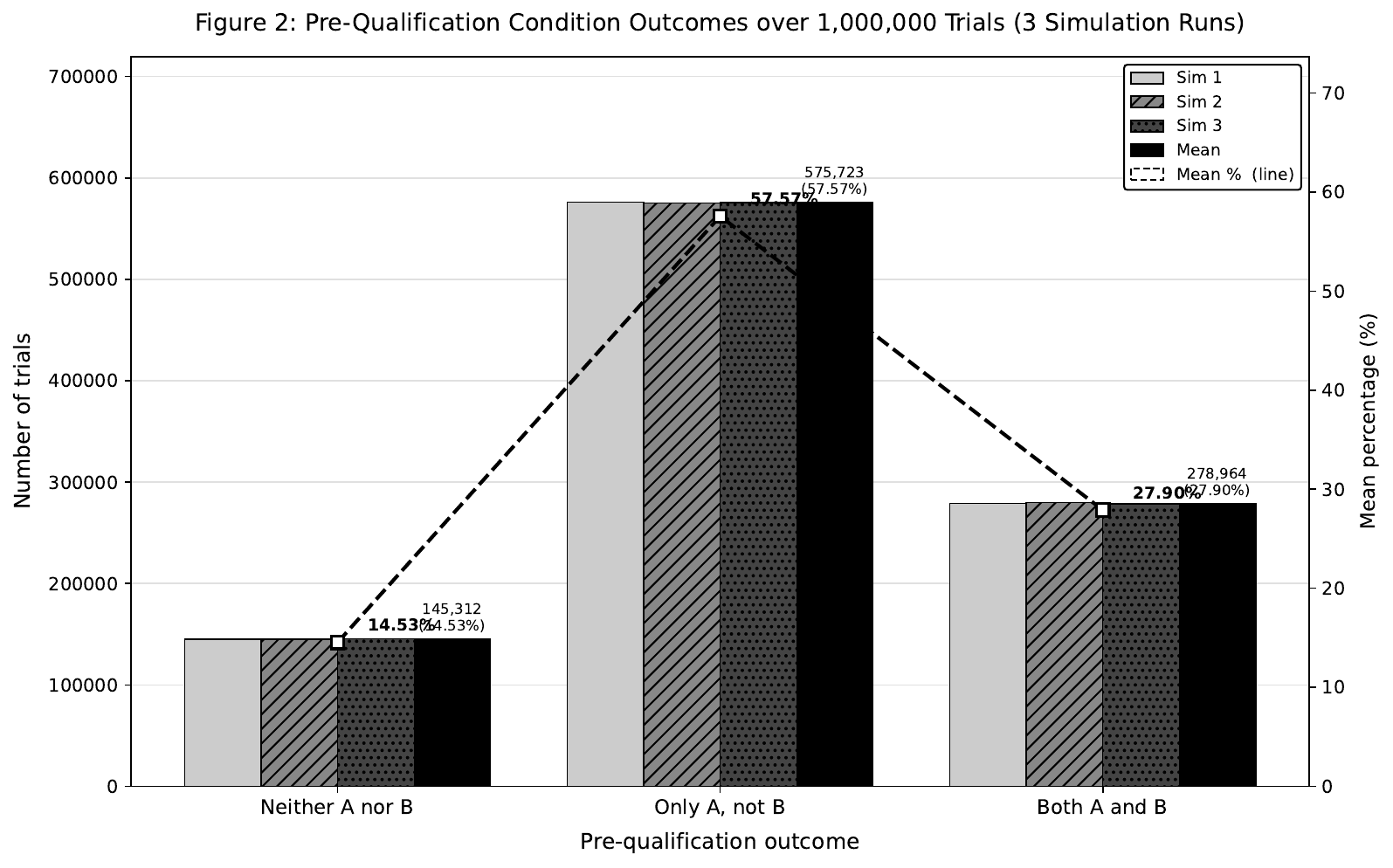}
    \caption{Pre-qualification condition outcomes over 1,000,000
             simulation trials across 3 independent runs.
             Grouped bars show trial counts per simulation run and
             mean (dark navy); the red line shows the mean percentage
             on the secondary axis.
             Approximately 14.53\% of deals failed Condition~A,
             57.57\% satisfied only Condition~A (not~B), and
             27.90\% satisfied both conditions and proceeded to the
             capture phase.}
    \label{fig:cond_outcomes}
\end{figure}

Figure~\ref{fig:cond_outcomes} visualises the pre-qualification
condition outcomes as a grouped bar chart with an overlaid percentage
line across the three simulation runs, with the mean shown in dark navy.
The dominance of the ``Only A, not~B'' category reflects the
combinatorial stringency of Condition~B: even when the global gap
budget is satisfied, the per-suit player count constraint frequently
fails for at least one suit.

\begin{figure*}[htbp]
    \centering
    \includegraphics[width=\textwidth]{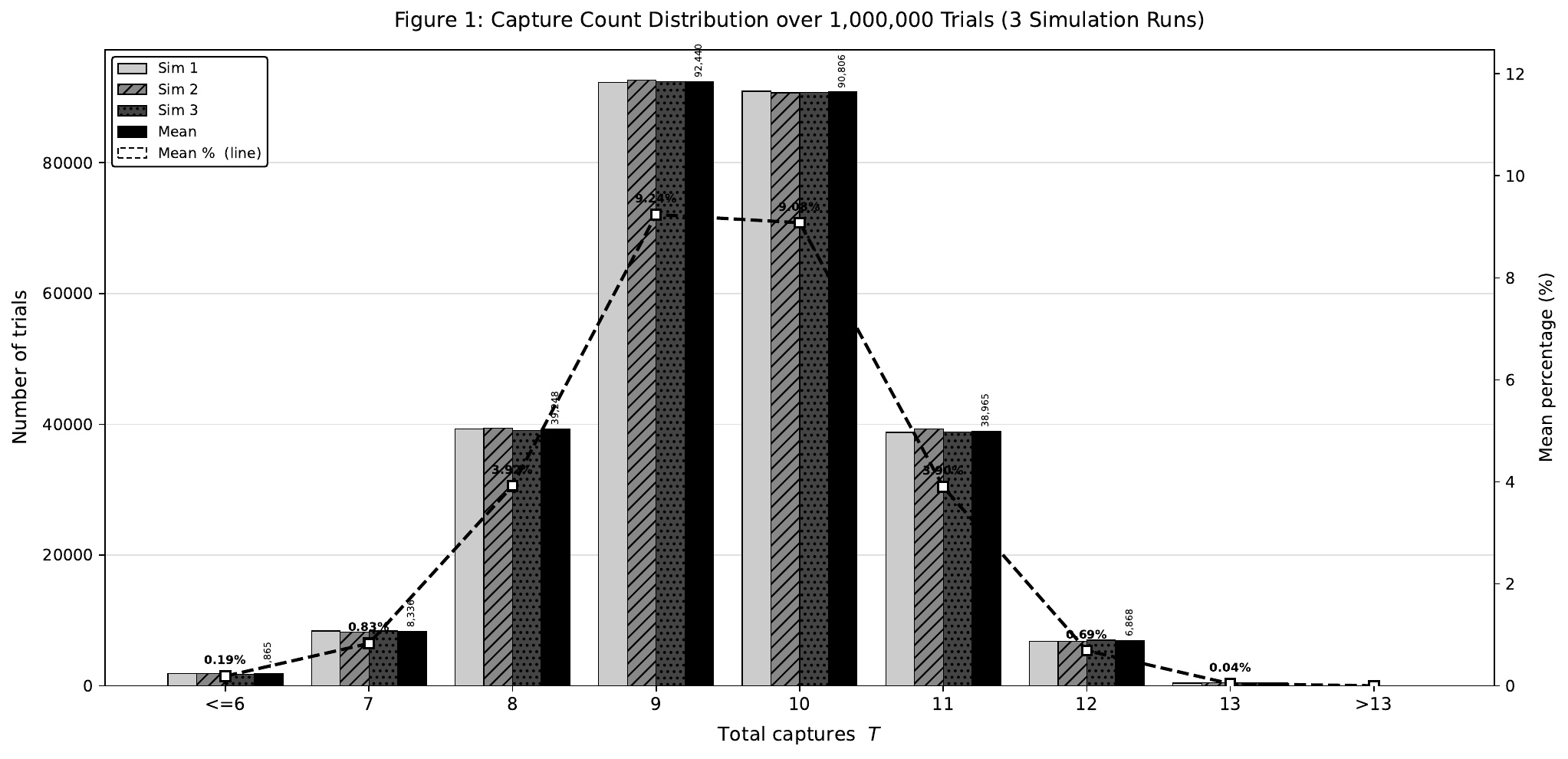}
    \caption{Distribution of total captures $T$ over 1,000,000
             simulation trials across 3 independent runs.
             Grouped bars show trial counts; the red line shows the
             mean percentage on the secondary axis.
             The distribution peaks at $T \in \{9, 10\}$, jointly
             accounting for approximately 65.6\% of all qualifying
             trials.
             No trial produced $T > 13$ across any of the three runs,
             validating the algorithm's theoretical stop condition.}
    \label{fig:capture_dist}
\end{figure*}

Figure~\ref{fig:capture_dist} presents the capture count distribution
as a grouped bar chart with an overlaid mean percentage line for all
qualifying trials.
The near-perfect overlap of the three runs across all bins confirms
that the distribution is stable and not sensitive to the random seed,
even at the scale of 1,000,000 trials per run.

\subsection{Discussion}

The simulation results at the scale of 1,000,000 trials per run
demonstrate three key properties of the algorithm.

First, the pre-qualification stage is a \emph{tight filter}: fewer
than one in three random deals (27.90\%) satisfies both Condition~A
and Condition~B and proceeds to the capture phase.
The dominant failure mode is Condition~B (57.57\% of all deals),
reflecting the combinatorial difficulty of simultaneously satisfying
the per-suit player count constraint across all four suits.

Second, among qualifying deals, the total capture count $T$ is
reliably concentrated in the range $T \in \{8, 9, 10, 11\}$,
accounting for over \textbf{93\%} of qualifying outcomes.
The distribution is stable across independent runs, confirming that
the greedy gap-minimisation strategy converges to consistent mid-range
capture totals regardless of random seed.

Third, the complete absence of $T > 13$ across all 3,000,000 combined
trials provides overwhelming empirical support for the correctness of
the algorithm's stop condition $\chi = T + (6 - Z) \geq 13$,
validating the algorithm's termination guarantee at large scale.

\section{Comparison Between 13-Card Single Hand Bidding Algorithm
vs Full Hand Final Output}

Conduct a comparative simulation study using \textbf{1,00,000
independent trials} to evaluate the difference between a player's
pre-bid trick estimate (based solely on his own 13 cards)(refer Sarkar et al.~\cite{b1}) and the
revised estimate after the partner's hand (dummy) is revealed.

\textbf{Algorithm~1} operates on a single hand of 13 cards
(13 ones, 39 zeros), representing the player's trick prediction
\emph{before} bidding --- without any knowledge of the partner's
holdings. This single-hand-only estimation problem closely parallels
the ``limit hand'' bid/no-bid classification problem studied via
inductive logic programming by Legras et al.~\cite{b10}, and the
supervised bidding-feature models of Amit and Markovitch~\cite{b7},
though our estimator uses closed-form structural conditions rather
than a learned classifier.

\textbf{Algorithm~2} operates on the combined 26-card layout
(13 ones, 13 twos, 26 zeros), representing the \emph{revised}
trick-winning potential computed after the dummy hand is opened
at the table.

\begin{figure}[htbp]
    \centering
    \includegraphics[width=\columnwidth]{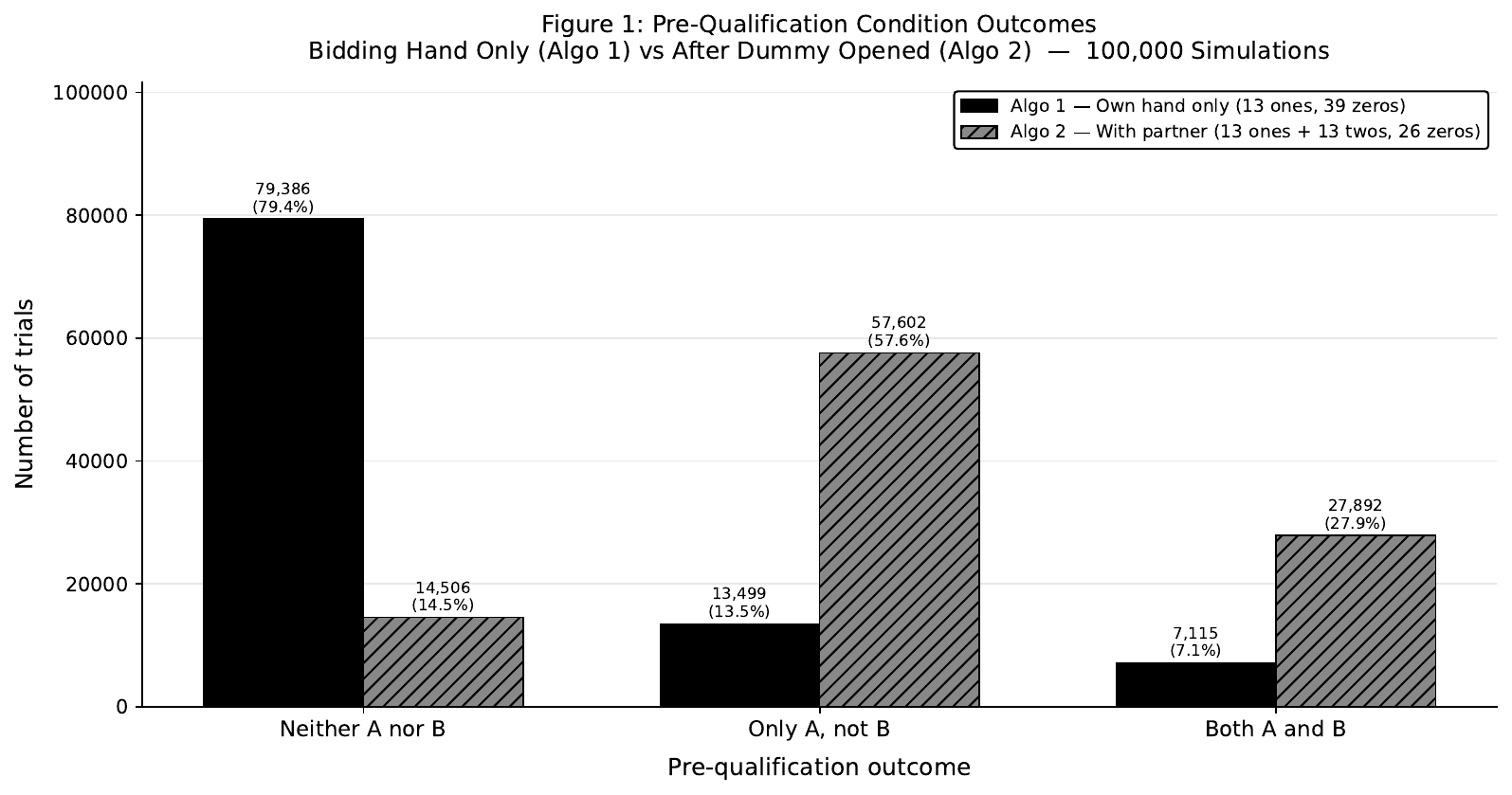}
    \caption{Pre-qualification condition outcomes: single hand
             bidding (Algo~1) vs full combined hand (Algo~2)
             over 1,00,000 simulations.}
    \label{fig:cond_compare}
\end{figure}

\begin{figure*}[htbp]
    \centering
    \includegraphics[width=\textwidth]{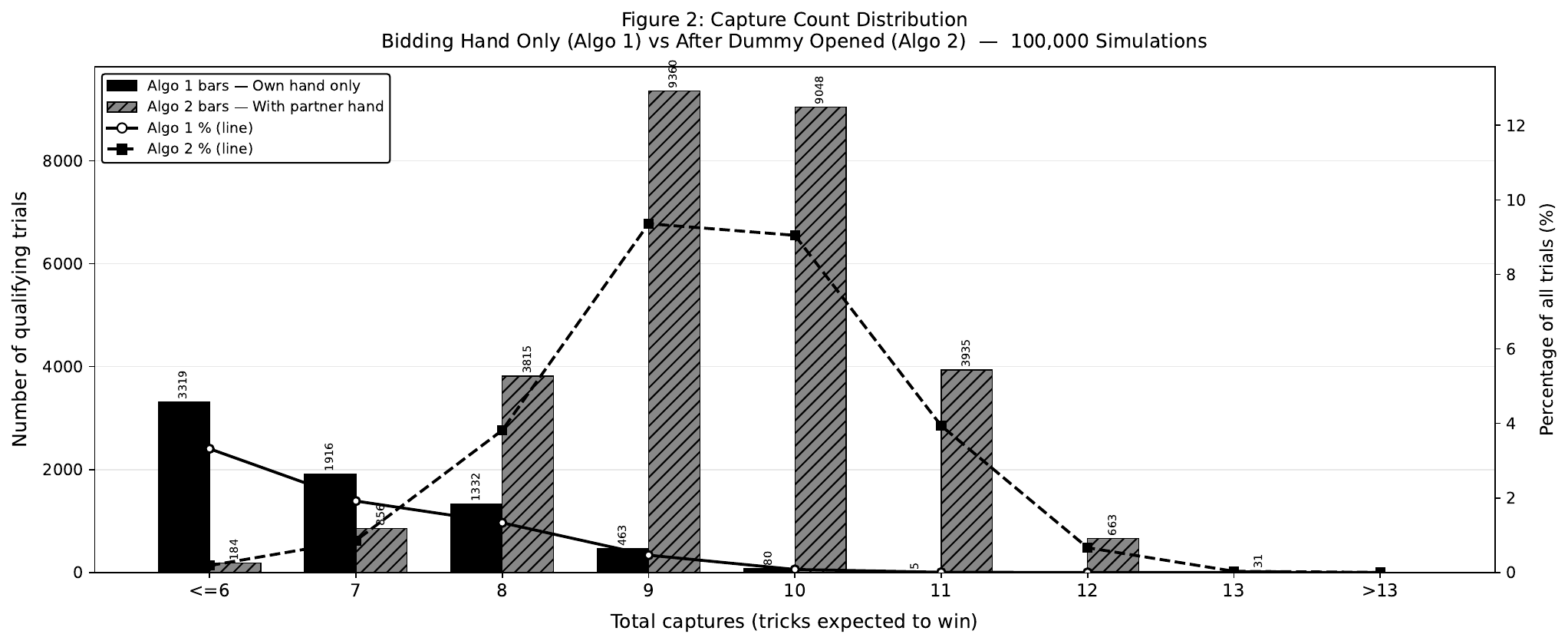}
    \caption{Capture count distribution: the shift between
             the pre-bid estimate (Algo~1) and the post-dummy
             revised output (Algo~2) over 1,00,000 simulations.}
    \label{fig:cap_compare}
\end{figure*}

\noindent The distributional shift between the two algorithms
quantifies precisely how much a player's trick expectation
changes once the dummy hand is revealed --- providing a
data-driven foundation for optimal bidding strategy in
Auction Bridge. This finding is consistent with the broader
observation, made across several independent bridge-AI research
efforts in different countries~\cite{b5,b8,b9}, that bidding
accuracy improves substantially once more of the deal's true
distribution becomes observable.

\section{Importance of 2nd Hand Criteria and Simulation Analysis}

The evaluation of combinatorial strategies in bridge-like card games reveals that success is not solely dependent on the primary hand's strength. While Algorithm 1 serves as an effective initial filter by measuring the ``Leading Gaps'' and card density of Player 1, the transition to the Global Capture Process (Algorithm 2) introduces a critical dependency on the second hand's distribution.

The ``Importance of 2nd Hand Criteria'' section highlights that the algorithm terminates immediately if the second hand fails to meet Condition B. Specifically, if Player 2's card count in any suit is less than or equal to the leading gap of that suit, the system deems the scenario unsustainable for a full capture sequence. As shown in the simulation data, approximately 25--28\% of cases that pass Algorithm 1 are subsequently rejected because of these second-hand deficiencies. This suggests that a successful global capture requires a synchronized distribution across both players to overcome the initial ``zero budget'' constraints, echoing the general finding in cooperative-partnership games that a partner's own hand quality --- not just the primary decision-maker's --- is often the binding constraint on joint performance~\cite{b8,b9}.

% Comparative table with adjusted column widths to prevent overflow
\begin{table}[htbp]
\caption{Comparative Analysis of Algorithm Satisfaction and Efficiency}
\label{tab:comparative}
\centering
\renewcommand{\arraystretch}{1.3}
\adjustbox{max width=\columnwidth}{%
\begin{tabular}{c c c c c c}
\toprule
\makecell{\textbf{Algo1}\\\textbf{Result}} &
\makecell{\textbf{A1}\\\textbf{Satisfied}} &
\makecell{\textbf{A2}\\\textbf{Satisfied}} &
\makecell{\textbf{A2 Not}\\\textbf{Satisfied}} &
\makecell{\textbf{P2}\\\textbf{Failure}} &
\makecell{\textbf{Success}\\\textbf{Rate (\%)}} \\
\midrule
7  & 893 & 641 & 252 & 252 & 71.78\% \\
8  & 544 & 410 & 134 & 134 & 75.37\% \\
9  & 228 & 176 & 52  & 52  & 77.19\% \\
10 & 93  & 69  & 24  & 24  & 74.19\% \\
\bottomrule
\end{tabular}%
}
\end{table}

\begin{figure}[htbp]
\centering
\includegraphics[width=\columnwidth]{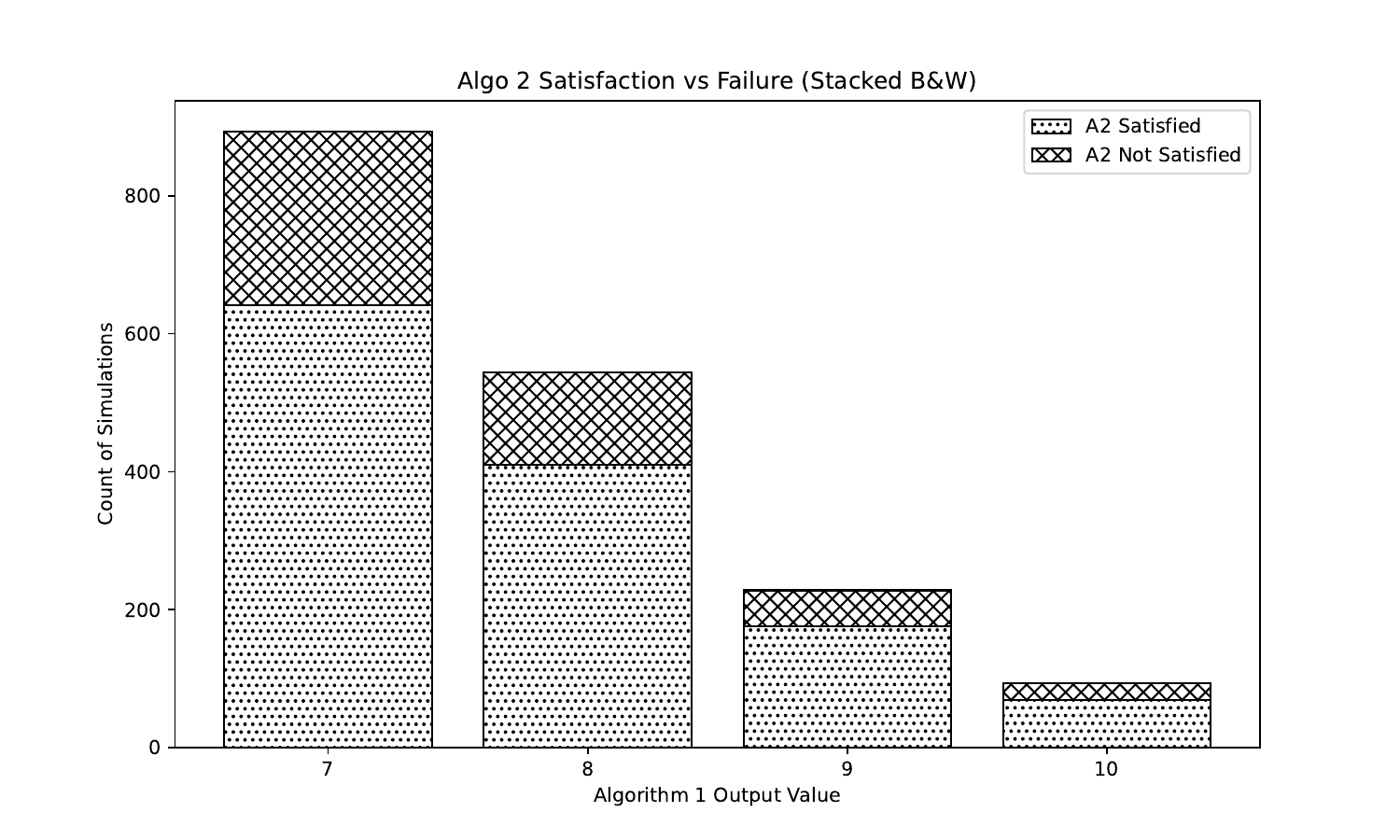}
\caption{Stacked black-and-white distribution showing the count of
         simulations where Algorithm 2 was satisfied versus cases
         missed due to second-hand criteria.}
\end{figure}

\section{Why We Ran the Same Simulation Three Times}
 
Any time a simulation depends on random shuffling, a single run only tells
part of the story. A single number could be a fluke -- a lucky (or unlucky)
sequence of shuffles that happens to push a statistic slightly higher or
lower than what the algorithm would produce on average. To make sure the
numbers we are reporting genuinely reflect how the bidding and play
algorithm behaves, rather than the particular shuffle order of one run, we
repeated the exact same 10{,}000{,}000-deal experiment three separate
times, each with a fresh, independent shuffling of the deck. If the three
runs agree closely with one another, we can be reasonably confident the
numbers are stable and trustworthy. If they disagree wildly, that would be
a warning sign that something in the simulation is unstable or that we need
many more deals before drawing conclusions.
 
The good news, as the rest of this section shows, is that the three runs
agree with each other almost to the decimal point.
 
\section{The Three Runs, Side by Side}
 
Table~\ref{tab:sim_summary} lays out the headline numbers from all three
runs. Each run shuffled and dealt 10 million hands, worked out who (if
anyone) could open a No-Trump bid under the qualification algorithm, played
out the resulting contract using the trick-capture algorithm, and scored
the result under both the old and the new scoring scheme.

\begin{table}[h]
\centering
\caption{Headline results from the three independent 10-million-deal runs.}
\label{tab:sim_summary}
\small
\setlength{\tabcolsep}{4pt}
\adjustbox{max width=0.92\columnwidth}{%
\begin{tabular}{lrrr}
\toprule
\textbf{Metric} & \textbf{Sim 1} & \textbf{Sim 2} & \textbf{Sim 3} \\
\midrule
Deals simulated                    & 10{,}000{,}000 & 10{,}000{,}000 & 10{,}000{,}000 \\
Deals with a valid bidder          & 1{,}030{,}367  & 1{,}029{,}511  & 1{,}028{,}714  \\
Deals where nobody could bid       & 8{,}969{,}633  & 8{,}970{,}489  & 8{,}971{,}286  \\
Overall win rate (\%)              & 65.86          & 66.00          & 66.03          \\
Losses caused by a weak dummy (\%) & 99.81          & 99.81          & 99.82          \\
Both bidder \& dummy eligible (n)  & 10{,}167       & 10{,}231       & 10{,}019       \\
Win rate when both eligible (\%)   & 93.42          & 93.50          & 93.75          \\
Avg.\ bidder level (both eligible) & 1.2102         & 1.2088         & 1.2188         \\
Avg.\ dummy level (both eligible)  & 1.2063         & 1.2081         & 1.2064         \\
Net points, old scheme             & 153{,}210      & 154{,}625      & 151{,}720      \\
Net points, new scheme             & 826{,}750      & 832{,}805      & 818{,}785      \\
\bottomrule
\end{tabular}%
}
\end{table}
 
\subsection*{The story the numbers tell}
 
Only around 10.3\% of all dealt hands ever produce a player whose own hand
is strong enough to open a No-Trump contract at all -- the remaining
roughly 89.7\% are simply passed out, with nobody's hand meeting the
qualification bar. That figure barely moves across the three runs (1.0304
million, 1.0295 million, and 1.0287 million qualifying deals out of 10
million), which is itself a small piece of evidence that the underlying
probability of a hand qualifying is a stable property of the algorithm,
not an artefact of any one shuffle sequence.
 
Once somebody does open a contract, the overall chance the contract is
actually made sits at almost exactly two-thirds (65.86\%, 66.00\%, and
66.03\% across the three runs) -- a spread of less than \emph{two-tenths
of one percentage point}. That is a remarkably tight cluster for a
random simulation of this size, and it tells us the win rate is a real,
repeatable property of the game, not noise.
 
The picture becomes even more interesting when we zoom in on the rare
cases where \emph{both} the winning bidder and the dummy independently
have hands strong enough to qualify on their own. This happens in roughly
0.1\% of all deals (around 10{,}000 to 10{,}200 out of 10 million each
time), but when it does happen, the win rate jumps sharply, from
two-thirds up into the low-to-mid 90s (93.42\%, 93.50\%, and 93.75\%).
In plain language: a hand that can open a bid on its own is a good sign,
but a partner who can \emph{also} open a bid on their own is a much
stronger sign that the partnership as a whole has the firepower to make
the contract.
 
We also see that virtually every loss (99.8\% or higher, in all three
runs) happens specifically because the dummy's hand did not independently
qualify. This is a consistent, repeatable finding across all three
simulations, and it lines up with the diagnostic work done earlier: the
capture algorithm itself is behaving correctly and resolving all thirteen
tricks in an orderly way; the shortfall comes from the fact that the
bidder's own hand strength says nothing about how strong their (unknown,
face-down) partner's hand happens to be.
 
Finally, the two scoring schemes tell two very different stories about
the same underlying games. Under the old scoring scheme, the net points
earned across all contracts (bidder's winnings minus the defenders'
undertrick penalties) sit in a tight band around 152--155 thousand points
across the three runs. Under the new scheme -- which rewards overtricks
more generously -- the same set of hands nets out to roughly 819--833
thousand points, more than five times higher. The relative ordering of
the three runs is consistent under both schemes (Sim 2 comes out
slightly ahead of Sim 1, which in turn is slightly ahead of Sim 3, for
the old scheme; a similar but not identical ordering holds for the new
scheme), underscoring that the choice of scoring rule changes the
\emph{scale} of the reward dramatically, without changing the underlying
gameplay it is rewarding.
 
\subsection*{A visual summary}
 
\begin{figure}[h]
\centering
\includegraphics[width=0.4\textwidth]{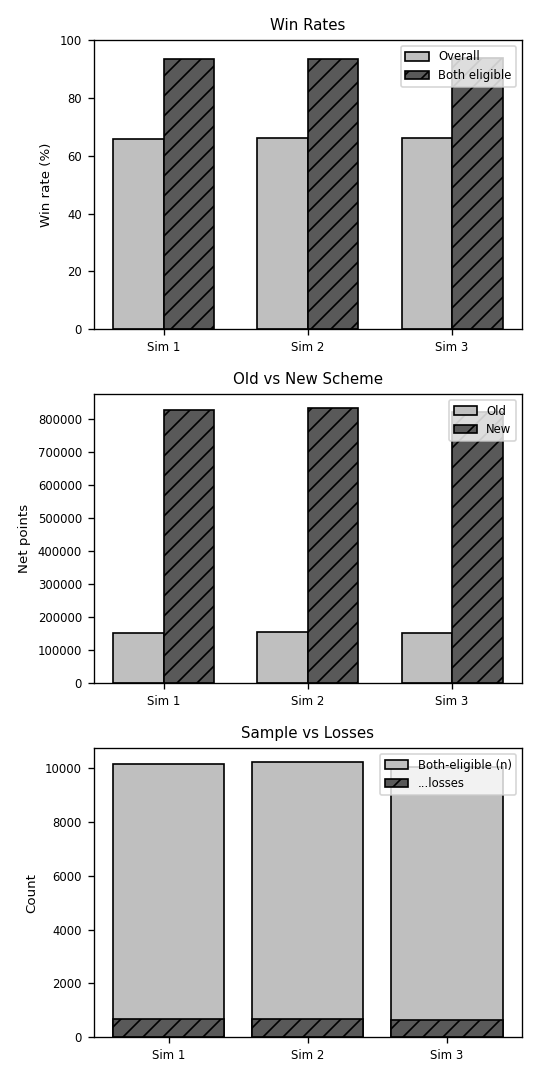}
\caption{Comparison of the three independent 10-million-deal simulation
runs. Left: overall win rate versus the win rate restricted to deals
where both the bidder and the dummy independently qualify to open a
contract. Middle: net points earned across all contracts, contrasting
the old and new scoring schemes. Right: the number of ``both eligible''
deals in each run, and how many of those ended in a loss.}
\label{fig:sim_comparison}
\end{figure}

Figure~\ref{fig:sim_comparison} puts these comparisons on a single page.
The left panel shows how tightly the overall win rate and the
both-eligible win rate cluster across the three runs. The middle panel
shows the dramatic difference in scale between the old and new scoring
schemes, while keeping the relative pattern across runs intact. The right
panel shows that although the sample of ``both eligible'' deals is small
relative to the full 10 million (roughly one in a thousand), it is
large enough in absolute terms (over ten thousand cases in each run) to
support a stable estimate of the win rate within that group.
  
\subsection*{Why the agreement between runs matters}
 
It would have been easy to run this simulation once, report the numbers,
and move on. Running it three times costs more computer time, but it
buys something valuable: confidence. A difference of less than 0.2
percentage points in the overall win rate, and less than 0.4 percentage
points in the harder-to-estimate ``both eligible'' win rate, across three
independent samples of ten million deals each, tells us the simulation
has converged. In other words, if we ran a fourth or a fifth batch of ten
million deals, we would expect the numbers to keep landing in this same
narrow range, rather than drifting off in a new direction. That is
exactly the kind of stability we want before drawing conclusions about
how the bidding and scoring rules actually behave.

\section{Conclusion}

This paper has presented a constant-time, structural alternative to
search- and learning-based approaches for estimating trick-taking
potential in a three-player auction bridge variant. Where prior
computational bridge research has largely relied on Monte Carlo
sampling~\cite{b5}, supervised or reinforcement learning over hand
features~\cite{b7,b8,b9}, or interpretable rule induction~\cite{b10},
our iterative capture algorithm derives an exact, reproducible trick
estimate directly from the structural properties of the dealt hand,
in time independent of the number of simulated trials. The broader
methodological toolkit of tree search~\cite{b13,b14,b6} and regret
minimisation~\cite{b11,b12} in other imperfect-information games
provides a useful frame of reference for future extensions of this
work, particularly for modelling the competitive bidding phase among
three simultaneously active players rather than the two-player
zero-sum capture setting studied here.

\subsection*{Where Reinforcement Learning Could Take This Next}

It is worth being upfront about what our algorithm is and is not. It
is a rule-based, closed-form estimator -- it does not learn anything,
and that is exactly why it runs in constant time. That is also its
main limitation: the greedy, smallest-gap-first capture rule behind
Algorithm~2 is a good heuristic against a randomly dealt hand, but
there is no reason to expect it is optimal against an opponent who is
actually trying to beat it. This is where reinforcement learning
becomes an interesting next step, not as a replacement for the
structural estimator, but as something built on top of it.

The most natural entry point is the bidding phase itself. Right now
we check feasibility against a single static snapshot of the hand, but
real bidding unfolds one bid at a time, and every bid an opponent
makes leaks a little information about what they are holding. That is
a textbook sequential decision problem: the state is your hand plus
whatever has been bid so far, the action is your next bid or a pass,
and the reward is however the round eventually scores under either of
the two schemes we compared in Section~VII. Yeh et al.~\cite{b8} and
Zhang et al.~\cite{b9} trained exactly this kind of bidding policy for
standard four-player bridge, but our three-player variant has three
independent bidders rather than fixed partnerships, so the multi-agent
setting is genuinely different and, as far as we can tell, largely
unexplored.

Card play after the dummy is revealed is the second obvious candidate.
Because we already frame that stage as a two-player zero-sum game in
Section~IV, it slots naturally into the same self-play training loop
that produced AlphaGo~\cite{b6} and that underlies the counterfactual
regret minimisation methods used in superhuman poker play~\cite{b11,
b12}. Training a play policy this way and then comparing it head-to-
head against our closed-form greedy rule would tell us something we
do not currently know: how much strategic ceiling the deterministic
algorithm is leaving on the table.

\subsection*{Where Bayesian Estimation Could Take This Next}

A second, complementary direction is to stop treating the opponent's
hand as a known quantity and start treating it as something we are
uncertain about and continually updating our beliefs on -- which is
really just a more principled version of what systems like
Ginsberg's GIB already do by sampling hidden hands~\cite{b5}.

The idea is straightforward. Before a card is played, all we know
about the opponent's hand is the prior implied by a random deal --
every combinatorially possible holding is equally likely. Every card
they play afterwards is evidence. If they fail to follow a suit, that
is about as strong a signal as you can get: it means they are holding
zero cards in it, and that constraint can be applied immediately. A
bid, similarly, is softer evidence about which suits they are likely
to be strong in, echoing the same feasibility logic already built into
Conditions A and B in Section~II. Bayes' rule ties this together:
\begin{multline}
  P(\text{opp.\ hand} \mid \text{evidence}) \;\propto\; \\
  P(\text{evidence} \mid \text{opp.\ hand}) \cdot P(\text{opp.\ hand}),
\end{multline}
and the posterior gets renormalised after every trick as new
information comes in. From there, the probability that a particular
card $c$ wins the current trick is just the posterior-weighted average
over all the hands the opponent could plausibly be holding,
\begin{multline}
  P(c \text{ wins}) \;=\; \sum_{h} P(h \mid \text{evidence so far}) \\
  \cdot\; \mathbb{1}\bigl[c \text{ beats best opposing card in } h\bigr].
\end{multline}
In practice this sum is far too large to compute exactly, so it would
be approximated the same way GIB approximates it~\cite{b5}: by
sampling a manageable set of plausible hands from the posterior and
averaging over those instead.

What this would mean for our algorithm specifically is that the
quantities we currently treat as fixed -- $c_1(s)$, $c_2(s)$, and
$\delta(s)$ -- would instead become random variables whose
distributions sharpen as the hand is played out, and the total capture
count $T$ would become an expected value rather than a single number
computed once at the start. The underlying structure of the algorithm
would not need to change; it would just be operating on beliefs
instead of certainties. And there is a nice synergy with the RL
direction above: an evolving posterior over the opponent's hand is
exactly the kind of belief state a learned bidding or play policy
would need as input, so the two extensions are not really separate
proposals so much as two pieces of the same longer-term system.

\newpage
%% --- References ---

\end{document}